# Quantum Logical Gates with Linear Quadripartite Cluster States of Continuous Variables

Aihong Tan    Changde Xie*    Kunchi Peng

(State Key Laboratory of Quantum Optics and Quantum Optics Devices, Institute of Opto-Electronics, Shanxi University, Taiyuan 030006, People's Republic of China )

**Abstract**

The concrete schemes to realize three types of basic quantum logical gates using linear quadripartite cluster states of optical continuous variables are proposed. The influences of noises and finite squeezing on the computation precision are analyzed in terms of the fidelity of propagated quantum information through the continuous variable cluster states. The proposed schemes provide direct references for the design of experimental systems implementing quantum computation with the cluster entanglement of amplitude and phase quadratures of light.

# 1 Introduction

Quantum computers (QC) promise efficient processing of certain computational tasks that are believed to be intractable with classical computer technology. Most of the concepts of quantum information and computation have been generalized in continuous variables (CV) [1] after they were initially developed for discrete variables (DV) [2]. A universal DV QC can perform any desired unitary transformation over discrete quantum variables by local operations, which are implemented on sequences of unitary quantum logic gates. Being different from the widely used quantum circuit model of QC [3], a novel model of quantum computation based on a highly entangled cluster state was proposed by Raussendorf and Briegel, in which the computation is completed only through single-qubit projective measurements [4]. Because of the essential role of measurement, the cluster based QC is irreversible, thus it was named the one-way QC [5]. The feasibility of one-way quantum computing has been experimentally demonstrated in single-photon regime with four-qubit cluster states [6-8].

In 1999, Lloyd and Braunstein provided necessary and sufficient conditions for constructing a universal CV QC and shown that QC over quadratures of the electromagnetic field might be realized using simple linear optical elements such as beam splitters and phase shifters, together with squeezers of light and nonlinear devices [9]. As a new type of multipartite entanglement, the conception of qubit-based cluster state was extended to CV and it was claimed that such states may be applied in quantum network communication but cannot be used in universal QC over CV because of their Gaussian character [10]. Successively, a universal QC model with CV cluster states was proposed by Menicucci et al. as a generalization of DV QC cluster-state model [11]. It was pointed out in Ref.[11] that the universal quantum computation based on CV cluster states can be performed only by adding to the toolbox (squeezed light, linear optics, and homodyne detection) any single-mode non-Gaussian measurement, while the initial cluster state itself remains Gaussian. In the proposed optical implementation of universal QC model using CV cluster states, squeezed-light sources serve as the nodes of the cluster, thus not only computation can be performed deterministically, but also the preparation of CV cluster states can be done unconditionally[11, 12]. Although the optical modes of the electromagnetic field provide a suitably experimental test bed for demonstrating the general principles of cluster-based QC, there is no any experimental result to be presented so far. We consider that the absence of the concrete design on the experimental systems is one of the reasons limited the progress of CV QC experimental research. Quantum logical gates are the most basic computing devices in QC which perform elementary quantum operations. To prompt the experimental study on QC with CV cluster states of light, we propose the schemes to realize the single-mode and multi-mode Gaussian quantum logical operations using linear quadripartite cluster states of electromagnetic field, which have been experimentally prepared [13, 14]. In Ref. [12], van Loock illustrated the principles of one-way QC using Gaussian CV cluster states with simple examples. Here, we will discuss concrete schemes for experimentally implementing quantum logical gates in one-way CV QC. The influences of the quantum noises and the finite squeezing of light on the computation precision will be analyzed in terms of fidelity of propagated quantum information through CV cluster states. Our analysis shows that finite squeezing reduces the precision of quantum logical operations. In practice, the ability of optical CV QC depends crucially on the squeezing degree of light used to prepare CV cluster states.

The paper is organized as follows: we simply describe the experimentally generating method of the quadripartite linear CV cluster states via the linear optical transformation of a pair of two-mode squeezed states of light produced from two non-degenerate optical amplifiers (NOPAs) in the second section. Then we introduce the schemes to realize the phase-space displacement transformation, the single-mode squeezing operation and the controlled-X operation using the cluster states in the sections 3 to 5, respectively. At last a brief conclusion is given in the section 6.

## 2 Preparation of quadripartite CV cluster states

The cluster state is a class of multipartite quantum entangled states and is classified in graph states. Originally, the term "cluster state" was introduced by Raussendorf and Briegel [15] to refer to the case where the graph $G$ is a two-dimensional square lattice states and they showed that the state can be used as a substrate for quantum computation. Generally, Graph quantum states are the multipartite entangled states that correspond to a certain mathematical graph $G$, i.e. a set of vertices connected by edges, where the vertices of the graph take the role of quantum systems and edges represent the physical interaction between the corresponding systems [16, 17]. CV cluster-like states proposed by J. Zhang and S. L. Braunstein [10] are a kind of CV Gaussian multipartite entangled states and the difference between CV cluster-like and GHZ-like states has been discussed in Ref. [10]. It has been pointed out that CV N-partite cluster-like states and GHZ-like states are not equivalent for $N > 3$, such as they have different persistence of entanglement and the criteria of quantum inseparability satisfied by them are also not the same [10, 13]. CV $N$-mode cluster state is a type of $N$-mode Gaussian states whose certain quadratures have perfect correlations in the limit of infinite squeezing, i.e. $\hat{Y}_a - \sum_{b \in N_a} \hat{X}_b \to 0$ ($a = 1\text{--}N$), where $Y_a$ and $X_b$ are quadrature phase and amplitude operators of optical modes $a$ and $b$ respectively, $N_a$ are the neighboring modes of $a$. The ideal CV cluster state is a simultaneous zero eigenstate of the quadrature combinations. Recently, it has been explicitly showed in Ref. [17] that there are different types of four-vertex graph states and all 4-mode CV cluster state graphs correspond either to a 4-mode GHZ(Greenberger-Horne-Zeilinger) entangled state or to a linear CV cluster state up to local Gaussian transformation and graph isomorphism. The difference and relationship between a variety of CV multipartite entangled states mirror the complexity of CV quantum systems. Although CV cluster states can be built deterministically, it will be impossible to create perfect CV cluster states due to the finite degree of squeezing obtainable in laboratories. The quantum entanglement of an experimentally generated cluster state should be verified by the sufficient condition for fully inseparability [18].

The imperfect CV four-mode cluster state of optical field has been experimentally prepared with the squeezed states of light and the linear optical transformation [13, 14]. The schematic diagram for the experimental generation of the four-mode linear CV cluster states, which will be used in following schemes for quantum logical gates, is shown in Fig.1. As that detailedly described in Ref. [13], two phase-quadrature squeezed states($a_1^s$, $a_4^s$) and two amplitude-quadrature squeezed states($a_2^s$, $a_3^s$) are simultaneously produced from a pair of NOPAs(NOPA1 and NOPA2), each of which consists of a type-II $\chi^2$ nonlinear optical crystal and an optical resonator [19]. The quadrature amplitudes ($X_{ai}$) and phases ($Y_{ai}$) of the four squeezed modes $a_i^s (i=1,2,3,4)$ equal to [20-22]

$$X_{a1(4)} = e^{+r} X_{a1(4)}^{(0)}, \qquad Y_{a1(4)} = e^{-r} Y_{a1(4)}^{(0)}, \qquad (1)$$

$$X_{a2(3)} = e^{-r} X_{a2(3)}^{(0)}, \qquad Y_{a2(3)} = e^{+r} Y_{a2(3)}^{(0)},$$

where, $X_{ai}^{(0)}$ and $Y_{ai}^{(0)}$ stand for the amplitude and the phase quadratures of the vacuum states ($a_i^{(0)}, i=1,2,3,4$) injected into NOPAs. The shot noise of a vacuum mode is normalized to 1. For simplification and without losing generality, we have assumed that the squeezing parameter of the four squeezed states is equal. The value of $r$ can be taken from zero to infinite, $r=0$ and $r=\infty$ correspond to no squeezing and perfect squeezing, respectively. The pump laser is a frequency-doubled CW laser, the output harmonic wave of which is used for the pump fields of the two NOPAs and the subharmonic wave serves as the injected signals($a_{01}, a_{02}, a_{03}$ and $a_{04}$) of the NOPAs as well as the local oscillators(LO) in the homodyne detections(see Fig.2 and 6). The beam splitters used in this system are chosen to completely eliminate all anti-squeezing components [14]. We take 1:4 beam splitter for BS$_1$, and 50% beam splitters for BS$_2$ and BS$_3$. At first interfering modes $a_2^s$ and $a_3^s$ on BS$_1$ with the phase difference of $\pi/2$ to produce two output modes $a_5$ and $a_6$, and then combining modes $a_1^s$ and $a_5$ on BS$_2$ with the phase difference of 0 and combining modes $a_4^s$ and $a_6$ on BS$_3$ with the phase difference of $\pi/2$, the final four output modes $b_i$ ($i=1,2,3,4$) are in a linear cluster state[10, 13, 14].

Based on Eq.(1), the combinations of the quadrature components ($X_{bi}$, $Y_{bi}$, $i=1,2,3,4$) of the four submodes in the cluster state with the squeezed noises can be expressed by the squeezing parameter $r$ of the original squeezed states [10, 13, 14]:

$$Y_{b1} - Y_{b2} = \sqrt{2}e^{-r}Y_{a1}^{(0)},$$

$$X_{b1} + X_{b2} + X_{b3} = \frac{\sqrt{10}}{2}e^{-r}X_{a2}^{(0)} - \frac{\sqrt{2}}{2}e^{-r}Y_{a4}^{(0)}, \qquad (2)$$

$$-Y_{b2} + Y_{b3} + Y_{b4} = -\frac{\sqrt{10}}{2}e^{-r}X_{a3}^{(0)} + \frac{\sqrt{2}}{2}e^{-r}Y_{a1}^{(0)},$$

$$X_{b3} - X_{b4} = -\sqrt{2}e^{-r}Y_{a4}^{(0)}.$$

It has been theoretically [10] and experimentally [13, 14] demonstrated that if the correlation variances of the amplitude quadratures($X_i$) and the phase quadratures($Y_i$) of the four modes $b_i$ satisfy the following inequalities, the four modes are in the quadripartite entangled linear cluster state with the full inseparability[18]:

$$\langle \delta^2(X_{b1} + X_{b2} + X_{b3}) \rangle + \langle \delta^2(Y_{b1} - Y_{b2}) \rangle < 4,$$

$$\langle \delta^2(X_{b3} - X_{b4}) \rangle + \langle \delta^2(-Y_{b2} + Y_{b3} + Y_{b4}) \rangle < 4, \qquad (3)$$

$$\langle \delta^2(X_{b1} + X_{b2} + X_{b3}) \rangle + \langle \delta^2(-Y_{b2} + Y_{b3} + Y_{b4}) \rangle < 4.$$

When all correlation combinations in the left-hand sides of these inequalities are smaller than the normalized shot noise limit of total four modes in the right-hand sides, the four optical modes $b_1 \sim b_4$ are in a cluster state with full quantum inseparability [10, 13, 14]. Substituting Eqs.(2) into the inequalities Eqs. (3), we can see that if the squeezing parameter $r$ is larger than a certain value, these inequalities will be met. The better squeezing (large $r$) corresponds to a better cluster state with higher quantum correlations of the quadrature combinations.

## 3 Single-mode evolution: Phase-space displacement operation

In CV regime, the Pauli $\hat{X}$ and $\hat{Z}$ operators are generalized to the Weyl-Heisenberg group, which is a Lie group with generators $\hat{q}$ and $\hat{p}$. The operators satisfy the canonical commutation relation $[\hat{q},\hat{p}]=i$ (with $\hbar = 1$). Then the $\sigma_x$ and $\sigma_z$ are generalized to the finite phase-space translation operators, $\hat{X}(s)=e^{-is\hat{p}}$ and $\hat{Z}(s)=e^{is\hat{q}}$ with $s \in R$ [11, 23]. As discussed in Ref. [11], the $\hat{Z}(s)=e^{is\hat{q}}$ gate is implemented by measuring $\hat{p}$ and subtracting $s$ from the result, where $s$ is the desired displacement.

The essence of cluster-state computation can be understood by considering a sequence of elementary teleportation circuits, in which the quantum information is transmitted through the cluster and potentially manipulated during each elementary step [3, 4]. In CV cluster-state quantum computation, the change of an initial quantum state during its propagation through the cluster depends on the choice of the measurement basis in each elementary step. As illustrated in Ref. [11, 12], the choice of the measurement basis corresponds to measurement $\hat{D}^+\hat{p}\hat{D}$, where $\hat{D}$ is an arbitrary operator diagonal in the computational basis (i. e., of the form $\exp[if(\hat{q})]$ ). Thus, the $\hat{Z}(s)=e^{is\hat{q}}$ gate is implemented by simply measuring $\hat{p}$ and subtracting $s$ from the result. The corresponding displacements will appear in the output state which can be corrected at the end. In experiments, the correction may conveniently be implemented with the amplitude and phase modulators.

In DV one-way computer, the known modification can be accounted for by adjusting the measurement basis for the final readout. But for a given finite size cluster, the output qubit may be the input qubit of subsequent circuit, so it must not be measured. To exam the operation result, the correction to the modification resulting from the measurement should be made on the output optic mode. Thus in the scheme of DV one-way computing, the AM and PM are used for the active feed-forward(see Ref. [7]). For the same reason, the use of AM and PM is also necessary in the CV scheme. The modulators are used to correct the corresponding displacements in the output state resulting from the measurement of the cluster state.

The experimental setup to realize the phase-space displacement operation is shown in Fig.2. Using the prepared quadripartite cluster states, we can choose arbitrarily two submodes, $b_1$ and $b_4$ for example, to be the input and output mode, respectively. In DV regime [4], any desired input state can be prepared by the other circuit preceding the proper circuit for computation; hence no input information needs to be written to the qubits before they are entangled. For the ideal case of quantum computation with perfect CV Cluster-state which is prepared by coupling perfect squeezed states, the perfect squeezed states are the eigenstates of a quadrature component, one of which may play the role of the input state. However, in experiments the produced cluster states are not able to be perfect and thus it is difficult to figure out an exact expression of the original squeezed state, which serves as the input state, from an imperfect cluster. For simplification and pedagogical reasons, in the discussion on the CV logical operation, we use the same method with Ref. [11, 12] where a cluster state is attached to a certain input state, which can be imaged as a part of another cluster state used in the preceding step, during the on-line computation.

The input state $a_{in}$ of the logical gate is combined with the mode $b_1$ at a 50:50 beam splitter with the phase difference of 0. In Heisenberg picture, the input state is an arbitrary Gaussian state and can be expressed as $a_{in} = X_{in} + iY_{in}$, $X_{in}$ and $Y_{in}$ are the amplitude and phase quadrature of $a_{in}$ respectively.

In Fig.2, the modes $c_1$ and $c_2$ with the amplitude quadratures $X_{c1}$, $X_{c2}$ and the phase quadratures $Y_{c1}$, $Y_{c2}$ are two output modes from a 50% beamsplitter, on which mode $a_{in}$ and mode $b_1$ are coupled with the phase difference of zero. The amplitude quadratures ($X_j$) and the phase quadratures ($Y_j$) ($j = c_1, c_2, b_2, b_3, b_4$) of modes $c_1$, $c_2$, $b_2$, $b_3$ and $b_4$ are expressed by

$$X_{c1} = (\frac{1}{\sqrt{5}} X_{a2} - \frac{1}{2\sqrt{5}} Y_{a3} + \frac{1}{2} X_{a1}) + \frac{1}{\sqrt{2}} X_{in},$$

$$Y_{c1} = (\frac{1}{\sqrt{5}} Y_{a2} + \frac{1}{2\sqrt{5}} X_{a3} + \frac{1}{2} Y_{a1}) + \frac{1}{\sqrt{2}} Y_{in},$$

$$X_{c2} = (\frac{1}{\sqrt{5}} X_{a2} - \frac{1}{2\sqrt{5}} Y_{a3} + \frac{1}{2} X_{a1}) - \frac{1}{\sqrt{2}} X_{in},$$

$$Y_{c2} = (\frac{1}{\sqrt{5}} Y_{a2} + \frac{1}{2\sqrt{5}} X_{a3} + \frac{1}{2} Y_{a1}) - \frac{1}{\sqrt{2}} Y_{in}, \qquad (4)$$

$$X_{b2} = \frac{2}{\sqrt{10}} X_{a2} - \frac{1}{\sqrt{10}} Y_{a3} - \frac{1}{\sqrt{2}} X_{a1}, \quad Y_{b2} = \frac{2}{\sqrt{10}} Y_{a2} + \frac{1}{\sqrt{10}} X_{a3} - \frac{1}{\sqrt{2}} Y_{a1},$$

$$X_{b3} = \frac{1}{\sqrt{10}} X_{a2} + \frac{2}{\sqrt{10}} Y_{a3} - \frac{1}{\sqrt{2}} Y_{a4}, \quad Y_{b3} = \frac{1}{\sqrt{10}} Y_{a2} - \frac{2}{\sqrt{10}} X_{a3} + \frac{1}{\sqrt{2}} X_{a4},$$

$$X_{b4} = \frac{1}{\sqrt{10}} X_{a2} + \frac{2}{\sqrt{10}} Y_{a3} + \frac{1}{\sqrt{2}} Y_{a4}, \quad Y_{b4} = \frac{1}{\sqrt{10}} Y_{a2} - \frac{2}{\sqrt{10}} X_{a3} - \frac{1}{\sqrt{2}} X_{a4}.$$

Where $X_{ai}$ and $Y_{ai}$ ($i = 1, 2, 3, 4$) are the amplitude and the phase quadratures of the initial squeezed states $a_i^s$ expressed in Eq.(1). At first, the amplitude and phase quadratures $X_{c1}$, $Y_{c2}$, $X_{b2}$ and $Y_{b3}$ are measured by the homodyne detectors HD$_o$ ($o = 1, 2, 3, 4$) respectively. The photocurrent of $X_{c1}(Y_{c2})$ measured by HD$_1$ (HD$_2$) is displaced an amount $s_0(s_1)$, which corresponding the desired displaced amount $s = \sqrt{2} s_0 (\sqrt{2} s_1)$. The sum of the photocurrent of the displaced $X_{c1}(Y_{c2})$ and the photocurrent of $X_{b2}(Y_{b3})$ measured by HD₃ (HD₄) are used to modulate the mode $b_4$ via an amplitude (phase) modulator AM (PM). The modulated mode $b_4$ is the resultant output mode $a^{out}$, the amplitude and phase quadratures of which are expressed by

$$X^{out} = X_{b4} + g_0(X_{c1} + s_0) + g_2 X_{b2}$$
$$= (\frac{1}{\sqrt{10}} X_{a2} + \frac{2}{\sqrt{10}} Y_{a3} + \frac{1}{\sqrt{2}} Y_{a4}) + g_0 \times [(\frac{1}{\sqrt{5}} X_{a2} - \frac{1}{2\sqrt{5}} Y_{a3} + \frac{1}{2} X_{a1}) + \frac{1}{\sqrt{2}} X_{in} + s_0]$$
$$+ g_2 \times (\frac{2}{\sqrt{10}} X_{a2} - \frac{1}{\sqrt{10}} Y_{a3} - \frac{1}{\sqrt{2}} X_{a1})$$
$$= (\frac{1}{\sqrt{10}} + \frac{g_0}{\sqrt{5}} + \frac{2g_2}{\sqrt{10}}) X_{a2} + (\frac{2}{\sqrt{10}} - \frac{g_0}{2\sqrt{5}} - \frac{g_2}{\sqrt{10}}) Y_{a3} + \frac{1}{\sqrt{2}} Y_{a4} + (\frac{g_0}{2} - \frac{g_2}{\sqrt{2}}) X_{a1}$$
$$+ \frac{g_0}{\sqrt{2}} X_{in} + g_0 s_0,$$

(5)

$$Y^{out} = Y_{b4} + g_1(Y_{c2} - s_1) + g_3 Y_{b3}$$
$$= (\frac{1}{\sqrt{10}} Y_{a2} - \frac{2}{\sqrt{10}} X_{a3} - \frac{1}{\sqrt{2}} X_{a4}) + g_1 \times [(\frac{1}{\sqrt{5}} Y_{a2} + \frac{1}{2\sqrt{5}} X_{a3} + \frac{1}{2} Y_{a1}) - \frac{1}{\sqrt{2}} Y_{in} - s_1]$$
$$+ g_3 \times (\frac{1}{\sqrt{10}} Y_{a2} - \frac{2}{\sqrt{10}} X_{a3} + \frac{1}{\sqrt{2}} X_{a4})$$

(6)

$$= (\frac{1}{\sqrt{10}} + \frac{g_1}{\sqrt{5}} + \frac{g_3}{\sqrt{10}}) Y_{a2} - (\frac{2}{\sqrt{10}} - \frac{g_1}{2\sqrt{5}} + \frac{2g_3}{\sqrt{10}}) X_{a3} + \frac{g_1}{2} Y a_1 + (-\frac{1}{\sqrt{2}} + \frac{g_3}{\sqrt{2}}) X_{a4}$$
$$- \frac{g_1}{\sqrt{2}} Y_{in} - g_1 s_0.$$

Where, $g_i$ $(i = 0,1,2,3)$ are the gain factors of the corresponding photocurrents and we take $g_0 = \sqrt{2}$, $g_1 = -\sqrt{2}$ to ensure the coefficient of $X_{in}$ and $Y_{in}$ in the output mode are 1. Substituting $g_0, g_1$ and Eq.(1) into Eqs.(5) and (6), we obtain

$$X^{out} = (\frac{3}{\sqrt{10}} + \frac{2g_2}{\sqrt{10}}) X_{a2} + (\frac{1}{\sqrt{10}} - \frac{g_2}{\sqrt{10}}) Y_{a3} + \frac{1}{\sqrt{2}} Y_{a4} + (\frac{1}{\sqrt{2}} - \frac{g_2}{\sqrt{2}}) X_{a1} + X_{in} + \sqrt{2} s_0, \quad (7)$$

$$Y^{out} = (-\frac{1}{\sqrt{10}} + \frac{g_3}{\sqrt{10}}) Y_{a2} - (\frac{3}{\sqrt{10}} + \frac{2g_3}{\sqrt{10}}) X_{a3} - \frac{1}{\sqrt{2}} Y a_1 + (-\frac{1}{\sqrt{2}} + \frac{g_3}{\sqrt{2}}) X_{a4} + Y_{in} + \sqrt{2} s_1. \quad (8)$$

The calculated fluctuation variances $\sigma_x^2$ and $\sigma_y^2$ of $X^{out}$ and $Y^{out}$ are

$$\sigma_x^2 = (\frac{3}{\sqrt{10}} + \frac{2g_2}{\sqrt{10}})^2 e^{-2r} + (\frac{1}{\sqrt{10}} - \frac{g_2}{\sqrt{10}})^2 e^{2r} + \frac{1}{2}e^{-2r} + (\frac{1}{\sqrt{2}} - \frac{g_2}{\sqrt{2}})^2 e^{2r} + V(X_{in}), \qquad (9)$$

$$\sigma_y^2 = (-\frac{1}{\sqrt{10}} + \frac{g_3}{\sqrt{10}})^2 e^{2r} + (\frac{3}{\sqrt{10}} + \frac{2g_3}{\sqrt{10}})^2 e^{-2r} + \frac{1}{2}e^{-2r} + (-\frac{1}{\sqrt{2}} + \frac{g_3}{\sqrt{2}})^2 e^{2r} + V(Y_{in}). \qquad (10)$$

Calculating the minimum values of $\sigma_x$ and $\sigma_y$ in terms of $g_2$ and $g_3$, we obtain the optimum gain factors

$$g_2^{opt} = g_3^{opt} = \frac{3(e^{2r} - e^{-2r})}{2e^{-2r} + 3e^{2r}}. \qquad (11)$$

The minimum variance equals

$$\sigma_{x,min}^2 = \frac{e^{-2r} + 9e^{2r}}{2 + 3e^{4r}} + V(X_{in}), \qquad (12)$$

$$\sigma_{y,min}^2 = \frac{e^{-2r} + 9e^{2r}}{2 + 3e^{4r}} + V(Y_{in}). \qquad (13)$$

From Eqs.(7) and (8), we can easily prove

$$\langle X^{out} \rangle = \langle X_{in} \rangle + \sqrt{2}s_0, \qquad \langle Y^{out} \rangle = \langle Y_{in} \rangle + \sqrt{2}s_1. \qquad (14)$$

Obviously, the average values of the amplitude and the phase quadratures of the input state have been displaced in the phase-space a desired amount $s = \sqrt{2}s_0$ and $s = \sqrt{2}s_1$, respectively.

According to the Rayleigh criterion in optics, when the center of the Airy disk for the first object occurs at the first minimum of the Airy disk of the second one, we say that the two objects can be barely resolved [24]. For a Gaussian wavepacket, it can be calculated based on the Rayleigh criterion that if taking $\delta x(\delta y) = 2\sigma_x$ and $\delta x(\delta y) = 3\sigma_x$, to be the radius of Airy disk (the first minimum), the resolving precision will reach 95% and 99%, respectively[24].

Thus we consider when

$$\sqrt{2}s_0 > 3\sigma_x = 3\times[\frac{e^{-2r}+9e^{2r}}{2+3e^{4r}}+V(X_{in})]^{\frac{1}{2}}, \tag{15}$$

$$\sqrt{2}s_1 > 3\sigma_y = 3\times[\frac{e^{-2r}+9e^{2r}}{2+3e^{4r}}+V(Y_{in})]^{\frac{1}{2}}, \tag{16}$$

the displacement in $x$ and $y$ direction can be distinguished. We define $\frac{3}{\sqrt{2}}\sigma_x$ and $\frac{3}{\sqrt{2}}\sigma_y$ to be the minimum of the displacement limited by the quantum noises in optical modes for a given $r$ and noises of the input state [$V(X_{in})$ and $V(Y_{in})$]. Only when the displacement $s_0(s_1)=s/\sqrt{2}$ is larger than the minimum, the displacement in the phase-space is distinguishable. The minimum distinguishable displacement $s_0^{\min}(s_1^{\min})$ stands for the reachable precision of a logical operation system.

For a general example, we assume that the input state is a squeezed state with a squeezing parameter of $r'$ ($r'=0$ corresponds to a coherent state). The dependences of the distinguishable displacements of the amplitude quadrature ($s_0$) and the phase quadrature ($s_1$) upon $r$ and $r'$ are shown in Fig.3. We can see that when $r$ and $r'$ increase, $s_0^{\min}$ and $s_1^{\min}$ decrease, however the influence of $r$ is lager than that of $r'$. It means that for performing a precise phase-space displacement operation on an input quantum state, we have to prepare a cluster state with high squeezing parameter at first.

When $s_0=0$ and $s_1=0$, the system performs an operation corresponding to an identity gate, in which the information propagates down a quantum wire to complete a simplest single-mode evolution. In fact, to propagate the information down a quantum wire, the basic method is teleportation [25-27]. Just like that in one-way DV QC scheme, a combination of successive one-qubit teleportation plays a key role [28, 29], CV teleportation is also the elementary method for performing CV quantum computation with cluster states. The identity operation is equivalent to the teleportation of the input state $a_{in}$ to the output state $a_{out}$ under the help of cluster entanglement. The flexibility of the system is that we can also extract the output state either from $b_2$ or $b_3$ instead of from $b_4$.

If using the unity gain ($g = 1$), the fidelity for the input Gaussian states is simply given by $F = \dfrac{2}{\sqrt{(1+\sigma_x^2)(1+\sigma_y^2)}}$ [26]. Substituting Eq.(12) and (13) into the fidelity formula, the dependence of $F$ on the squeezing parameter $r$ in the system is shown in Fig. 4.

For perfect initial squeezing of $r \to \infty$, the fidelity $F \to 1$, it means that in the ideal case the quantum information is successfully propagated down the quantum wire. Generally, for the classical case without squeezing ($r \to 0$), the best fidelity $F$ should equal 0.5[26], which just is the result in Fig.4.

## 4 Single-mode squeezing operation

A single-mode squeezer is an important primitive for performing Gaussian transformation. As pointed out in Ref. [11, 12], in a squeezer there is the operator of quadratic form, $D = \exp(it\hat{q}^2)$, which can be performed via a given cluster state solely by doing suitable homodyne measurements, where $t$ stands for the squeezing parameter of the $D = \exp(it\hat{q}^2)$ operation. The experimental setup of the single-mode squeezer is the same as Fig.2. However in the squeezing operation, a linear combination of position and momentum should be detected with the homodyne detections (HDs), which correspond to the measurement of rotated quadratures[12].

Coupling the input state $a_{in}$ to a submode $b_1$ of the quadripartite cluster state,

and adjusting the phase differences between the local oscillator and the signal field in HD$_1$, HD$_2$, HD$_3$ and HD$_4$ to $\theta$, $\pi/2$, $0$ and $\pi/2$ for measuring $Y_{c1}\sin\theta + X_{c1}\cos\theta$, $Y_{c2}$, $X_{b2}$, and $Y_{b3}$, respectively. Then, those measured photocurrents are used for displacing the amplitude and the phase quadratures of the mode $b_4$. The quadratures of the output mode are expressed by

$$X^{out} = X_{b4} + \sqrt{2}\frac{1}{\cos\theta}(\cos\theta X_{c1} + \sin\theta Y_{c1}) + X_{b2} - \sqrt{2}\tan\theta Y_{c2}$$

$$= (\frac{1}{\sqrt{10}}X_{a2} + \frac{2}{\sqrt{10}}Y_{a3} + \frac{1}{\sqrt{2}}Y_{a4}) + \sqrt{2}\times\{[(\frac{1}{\sqrt{5}}X_{a2} - \frac{1}{2\sqrt{5}}Y_{a3} + \frac{1}{2}X_{a1}) + \frac{X_{in}}{\sqrt{2}}]$$

$$+ \tan\theta[(\frac{1}{\sqrt{5}}Y_{a2} + \frac{1}{2\sqrt{5}}X_{a3} + \frac{1}{2}Y_{a1}) + \frac{1}{\sqrt{2}}Y_{in}]\} + (\frac{2}{\sqrt{10}}X_{a2} - \frac{1}{\sqrt{10}}Y_{a3} - \frac{1}{\sqrt{2}}X_{a1}) \quad (17)$$

$$-\sqrt{2}\tan\theta[(\frac{1}{\sqrt{5}}Y_{a2} + \frac{1}{2\sqrt{5}}X_{a3} + \frac{1}{2}Y_{a1}) - \frac{1}{\sqrt{2}}Y_{in}]$$

$$= \sqrt{\frac{5}{2}}X_{a2} + \frac{1}{\sqrt{2}}Y_{a4} + X_{in} + 2\tan\theta Y_{in},$$

$$Y^{out} = Y_{b4} - \sqrt{2}\times Y_{c2} + Y_{b3}$$

$$= (\frac{1}{\sqrt{10}}Y_{a2} - \frac{2}{\sqrt{10}}X_{a3} - \frac{1}{\sqrt{2}}X_{a4}) - \sqrt{2}\times[(\frac{1}{\sqrt{5}}Y_{a2} + \frac{1}{2\sqrt{5}}X_{a3} + \frac{1}{2}Y_{a1}) - \frac{1}{\sqrt{2}}Y_{in})$$

$$+ (\frac{1}{\sqrt{10}}Y_{a2} - \frac{2}{\sqrt{10}}X_{a3} + \frac{1}{\sqrt{2}}X_{a4}) \quad (18)$$

$$= -\sqrt{\frac{5}{2}}X_{a3} - \frac{1}{\sqrt{2}}Ya_1 + Y_{in}.$$

In the equation (17), the rescaling factor is $\cos\theta$, and the squeezing parameter $t = -\tan\theta$. For experiments, the squeezing of the output mode $a^{out}$ can be checked with another homodyne detection HD$_5$. If the phase difference between the LO and $a_{out}$ in HD$_5$ is $\phi$, we have

$$Y^{out}\sin\phi + X^{out}\cos\phi$$

$$=\left[-\sqrt{\frac{5}{2}}X_{a3}-\frac{1}{\sqrt{2}}Ya_1+Y_{in}\right]\sin\phi+\left[\sqrt{\frac{5}{2}}X_{a2}+\frac{1}{\sqrt{2}}Y_{a4}+X_{in}-2\tan\theta Y_{in}\right]\cos\phi$$

$$=\left[-\sqrt{\frac{5}{2}}X_{a3}-\frac{1}{\sqrt{2}}Ya_1\right]\sin\phi+\left[\sqrt{\frac{5}{2}}X_{a2}+\frac{1}{\sqrt{2}}Y_{a4}\right]\cos\phi \tag{19}$$

$$+\cos\phi X_{in}+(2\tan\theta\cos\phi+\sin\phi)Y_{in},$$

$$V(Y^{out}\sin\phi + X^{out}\cos\phi) = 3e^{-2r} + \cos^2\phi + (2\tan\theta\cos\phi+\sin\phi)^2. \tag{20}$$

From Eq. (20), we can see that the fluctuation variances may be smaller than the normalized shot noise limit (SNL) for appropriate $\theta$ and $\phi$. The dependences of the variances $V$ in Eq. (20) on the detection phase $\phi$ of the output mode are drawn for different $\theta$ and a given $r=2$ in Fig.5. Obviously, the noise ellipse of the output squeezed mode becomes more narrow and the lowest variance becomes smaller (squeezing increases) when $\theta$ increases, which corresponds to the result in Ref. [11]. However, if $\tan\theta = 0$, we have $V(Y^{out}\sin\phi + X^{out}\cos\phi) = 3e^{-2r}+1$. In this case the variance $V$ does not depend on $\phi$, thus there is no squeezing to be generated whatever cluster is applied. In Fig.6 the functions of $V$ vs $\phi$ for different $r$ of the initial cluster state and a given $\tan\theta = 2$ are presented. It is pointed out that only when $r$ of the cluster state is larger than a threshold ($r=0.55$ in this example) squeezing of the output mode exists, i.e. $V$ is lower than the normalized SNL. Where $r=0.6$ and $r=1.15$ correspond to the squeezing of 5.2 dB and 10 dB respectively, which have been experimentally realized [30-33]. The maximum squeezing direction $\phi$ depends on $\theta$ only and does not on $r$. The dependence of $\phi^{opt}$ for the minimum $V_{min}$ on $\theta$ is expressed in Eq.(21):

$$\tan 2\phi^{opt} = (\tan\theta)^{-1} \tag{21}$$

The minimum $V_{min}$ for a given $r$ equals:

$$V(Y^{out}\cos\phi^{opt} + X^{out}\sin\phi^{opt}) = 3e^{-2r} + (\tan\phi^{opt})^{-2} \qquad (22)$$

## 5 Controlled-X operation

After the pauli $\hat{X}$ and $\hat{Z}$ operators are generalized to the finite phase-space translation operators, the CNOT and CPHASE are naturally generalized to controlled-$\hat{X}$ ($\hat{C}_X$) and controlled-$\hat{Z}$ ($\hat{C}_Z$), respectively, which effect a phase-space displacement on the target by an amount determined by the position eigenvalue of the control state: $\hat{C}_X = \exp(-i\hat{q}\otimes\hat{p})$ and $\hat{C}_Z = \exp(i\hat{q}\otimes\hat{q})$, where the order of the system is (control $\otimes$ target)[11]. In this section, we will discuss the realization of controlled-X operation.

Fig.7 is the proposed experimental scheme for realizing CV controlled-X operation using linear quadripartite Cluster state. The control signal $a_c$ and the target signal $a_t$ are expressed by:

$$a_c = X_c + iY_c, \qquad (23)$$

$$a_t = X_t + iY_t. \qquad (24)$$

Where $X_{c(t)}$ and $Y_{c(t)}$ are the amplitude and the phase quadrature of $a_{c(t)}$, $\langle X_{c(t)}\rangle = s_{c(t)}$, $s_c$ and $s_t$ stand for the position displacements of the control and the target signals in the phase-space related the zero point, respectively. The input control signal $a_c$ and the target signal $a_t$ are coupled respectively to the submodes $b_3$ and $b_2$ of the cluster state at a 50:50 beam splitter with the phase difference of 0. The quadratures of the coupled state equal to

$$X_{b1} = \frac{2}{\sqrt{10}}X_{a2} - \frac{1}{\sqrt{10}}Y_{a3} + \frac{1}{\sqrt{2}}X_{a1}, \qquad Y_{b1} = \frac{2}{\sqrt{10}}Y_{a2} + \frac{1}{\sqrt{10}}X_{a3} + \frac{1}{\sqrt{2}}Y_{a1},$$

$$X_{t1} = \frac{1}{\sqrt{2}}(\frac{2}{\sqrt{10}}X_{a2} - \frac{1}{\sqrt{10}}Y_{a3} - \frac{1}{\sqrt{2}}X_{a1} + X_t),$$

$$Y_{t1} = \frac{1}{\sqrt{2}}(\frac{2}{\sqrt{10}}Y_{a2} + \frac{1}{\sqrt{10}}X_{a3} - \frac{1}{\sqrt{2}}Y_{a1} + Y_t),$$

$$X_{t2} = \frac{1}{\sqrt{2}}(\frac{2}{\sqrt{10}}X_{a2} - \frac{1}{\sqrt{10}}Y_{a3} - \frac{1}{\sqrt{2}}X_{a1} - X_t),$$

$$Y_{t2} = \frac{1}{\sqrt{2}}(\frac{2}{\sqrt{10}}Y_{a2} + \frac{1}{\sqrt{10}}X_{a3} - \frac{1}{\sqrt{2}}Y_{a1} - Y_t),$$

$$X_{c1} = \frac{1}{\sqrt{2}}(\frac{1}{\sqrt{10}}X_{a2} + \frac{2}{\sqrt{10}}Y_{a3} - \frac{1}{\sqrt{2}}Y_{a4} - X_c), \qquad (25)$$

$$Y_{c1} = \frac{1}{\sqrt{2}}(\frac{1}{\sqrt{10}}Y_{a2} - \frac{2}{\sqrt{10}}X_{a3} + \frac{1}{\sqrt{2}}X_{a4} - Y_c),$$

$$X_{c2} = \frac{1}{\sqrt{2}}(\frac{1}{\sqrt{10}}X_{a2} + \frac{2}{\sqrt{10}}Y_{a3} - \frac{1}{\sqrt{2}}Y_{a4} + X_c),$$

$$Y_{c2} = \frac{1}{\sqrt{2}}(\frac{1}{\sqrt{10}}Y_{a2} - \frac{2}{\sqrt{10}}X_{a3} + \frac{1}{\sqrt{2}}X_{a4} + Y_c),$$

$$X_{b4} = \frac{1}{\sqrt{10}}X_{a2} + \frac{2}{\sqrt{10}}Y_{a3} + \frac{1}{\sqrt{2}}Y_{a4}, \qquad Y_{b4} = \frac{1}{\sqrt{10}}Y_{a2} - \frac{2}{\sqrt{10}}X_{a3} - \frac{1}{\sqrt{2}}X_{a4}.$$

Measuring $X_{t1}$, $Y_{t2}$, $X_{c1}$, $Y_{c2}$, and feeding forward the measured photocurrents to mode $b_1$ and $b_4$ respectively, the quadratures of the output mode become

$$\begin{aligned}X_t^{out} &= X_{b1} + \sqrt{2}X_{t1} + \sqrt{2}X_{c1} \\ &= (\frac{2}{\sqrt{10}}X_{a2} - \frac{1}{\sqrt{10}}Y_{a3} + \frac{1}{\sqrt{2}}X_{a1}) + \sqrt{2} \times \frac{1}{\sqrt{2}}(\frac{2}{\sqrt{10}}X_{a2} - \frac{1}{\sqrt{10}}Y_{a3} - \frac{1}{\sqrt{2}}X_{a1} + X_t) \\ &\quad + \sqrt{2} \times \frac{1}{\sqrt{2}}(\frac{1}{\sqrt{10}}X_{a2} + \frac{2}{\sqrt{10}}Y_{a3} - \frac{1}{\sqrt{2}}Y_{a4} - X_c) \\ &= \sqrt{\frac{5}{2}}X_{a2} - \sqrt{\frac{1}{2}}Y_{a4} + X_t - X_c,\end{aligned}$$

$$\begin{aligned}
Y_c^{out} &= Y_{b1} - \sqrt{2}Y_{t2} \\
&= \frac{2}{\sqrt{10}}Y_{a2} + \frac{1}{\sqrt{10}}X_{a3} + \frac{1}{\sqrt{2}}Y_{a1} - \sqrt{2} \times \frac{1}{\sqrt{2}}(\frac{2}{\sqrt{10}}Y_{a2} + \frac{1}{\sqrt{10}}X_{a3} - \frac{1}{\sqrt{2}}Y_{a1} - Y_t) \quad (26) \\
&= \sqrt{2}Y_{a1} + Y_t,
\end{aligned}$$

$$\begin{aligned}
X_c^{out} &= X_{b4} - \sqrt{2}X_{c1} \\
&= (\frac{1}{\sqrt{10}}X_{a2} + \frac{2}{\sqrt{10}}Y_{a3} + \frac{1}{\sqrt{2}}Y_{a4}) - \sqrt{2} \times \frac{1}{\sqrt{2}}(\frac{1}{\sqrt{10}}X_{a2} + \frac{2}{\sqrt{10}}Y_{a3} - \frac{1}{\sqrt{2}}Y_{a4} - X_c) \\
&= \sqrt{2}Y_{a4} + X_c,
\end{aligned}$$

$$\begin{aligned}
Y_t^{out} &= Y_{b4} - \sqrt{2}Y_{t2} + \sqrt{2}Y_{c2} \\
&= \frac{1}{\sqrt{10}}Y_{a2} - \frac{2}{\sqrt{10}}X_{a3} - \frac{1}{\sqrt{2}}X_{a4} - \sqrt{2} \times \frac{1}{\sqrt{2}}(\frac{2}{\sqrt{10}}Y_{a2} + \frac{1}{\sqrt{10}}X_{a3} - \frac{1}{\sqrt{2}}Y_{a1} - Y_t) \\
&\quad + \sqrt{2} \times \frac{1}{\sqrt{2}}(\frac{1}{\sqrt{10}}Y_{a2} - \frac{2}{\sqrt{10}}X_{a3} + \frac{1}{\sqrt{2}}X_{a4} + Y_c) \quad (27) \\
&= -\sqrt{\frac{5}{2}}X_{a3} + \sqrt{\frac{1}{2}}Y_{a1} + Y_t + Y_c.
\end{aligned}$$

The average values and the variances of the amplitude and the phase quadratures for the input and the output states are listed in table 1. We can see, the phase quadrature of the output target signal has been displaced under the control of the control signal $s_c$. It means that the controlled-X operation has been implemented.

Table 1 The average values and the variances of the amplitude and the phase quadratures for the input and the output states

|  | Control | Target |
|---|---|---|
| Input signal | $\langle X_c \rangle = s_c$    $\sigma_x = V(X_c)$ <br> $\langle Y_c \rangle = 0$    $\sigma_y = V(Y_c)$ | $\langle X_t \rangle = s_t$    $\sigma_x = V(X_t)$ <br> $\langle Y_t \rangle = 0$    $\sigma_y = V(Y_t)$ |
| Output signal | $\langle X_c^{out} \rangle = s_c$ <br> $\langle Y_c^{out} \rangle = 0$ <br> $\sigma_x = 2e^{-2r} + V(X_c) \xrightarrow{r \to \infty} V(X_c)$ <br> $\sigma_y = 3e^{-2r} + [V(Y_c) + V(Y_t)]$ <br> $\xrightarrow{r \to \infty} [V(Y_c) + V(Y_t)]$ | $\langle X_t^{out} \rangle = s_t - s_c$ <br> $\langle Y_t^{out} \rangle = 0$ <br> $\sigma_x = 3e^{-2r} + V(X_c) + V(X_t)$ <br> $\xrightarrow{r \to \infty} V(X_c) + V(X_t)$ <br> $\sigma_x = 2e^{-2r} + V(Y_t) \xrightarrow{r \to \infty} V(Y_t)$ |

For clearly exhibiting the effect of finite squeezing of the cluster state on the feature of output states, the Wigner functions of the input(output) control and target signals are shown in Fig.8, where we have assumed that both input control and target signals are the amplitude-squeezed states of light, and $s_c = 1$ and $s_t = 2$ (normalized to the shot noise limit). Obviously, the amplitude quadratures are displaced an amount along the direction of $x$ axis under the action of the control signal (from 2 to 1). Since the finite squeezing of the cluster state, some noises are added in the process, and thus the Wigner functions of the output states are expanded at the direction of $x$ axis. It means that the imperfect cluster will result in the squeezing decrease of input state. The influence will reduce when the squeezing parameter $r$ of the cluster state increases (comparing $r = 1$ and $r = 3$).

# 6 Conclusions

For conclusion, following the theoretical suggestions on CV QC in Ref.[11] and [12], we designed the concrete experimental systems for implementing the phase-space displacement transformation, squeezing and controlled-X operation based on the linear quadripartite cluster state of electromagnetic field. In the proposed schemes only linear optics, homodyne detections and classical feedforwards are required and the cluster state can be prepared off-line. The influences of finite squeezing of cluster state on the precisions of the logical operations are analyzed. Although a nonlinear element such as any single-mode non-Gaussian measurement is needed for demonstrating universal CV QC, the realization of the proposed logical operations is the first step for universal quantum computation. The calculations and discussions in this paper provide direct references for the design of the experimental systems implementing CV logical gates. The linear CV quadripartite cluster states have been experimentally obtained [13, 14], thus the proposed schemes for the CV logical operations are accessible with the present experimental technology. In the presented paper, we only analyzed the Gaussian optical modes and the analyses based on the quantum variances can not be applied in the non-Gaussian states of the optical field. The feasible scheme for the quantum computation using non-Gaussian optical states still keeps being an open question.

This research was supported by Natural Science Foundation of China (Grants No. 60736040 and 10674088), National Basic Research Program of China(Grant No.2006CB921101), the PCSIRT (Grant No. IRT0516).

* Corresponding author's Email address: changde@sxu.edu.cn

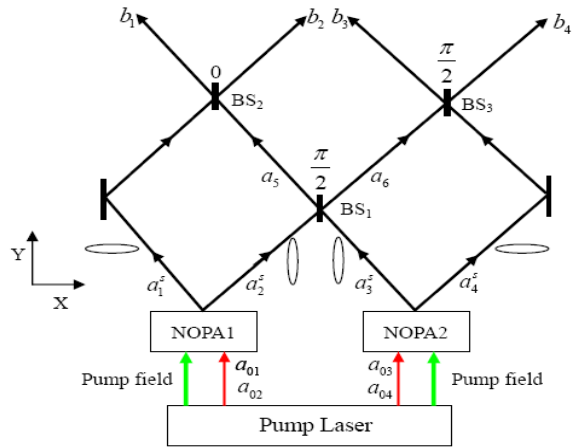

Fig.1 (Color online) Principle schematic for CV quadripartite linear cluster state generation

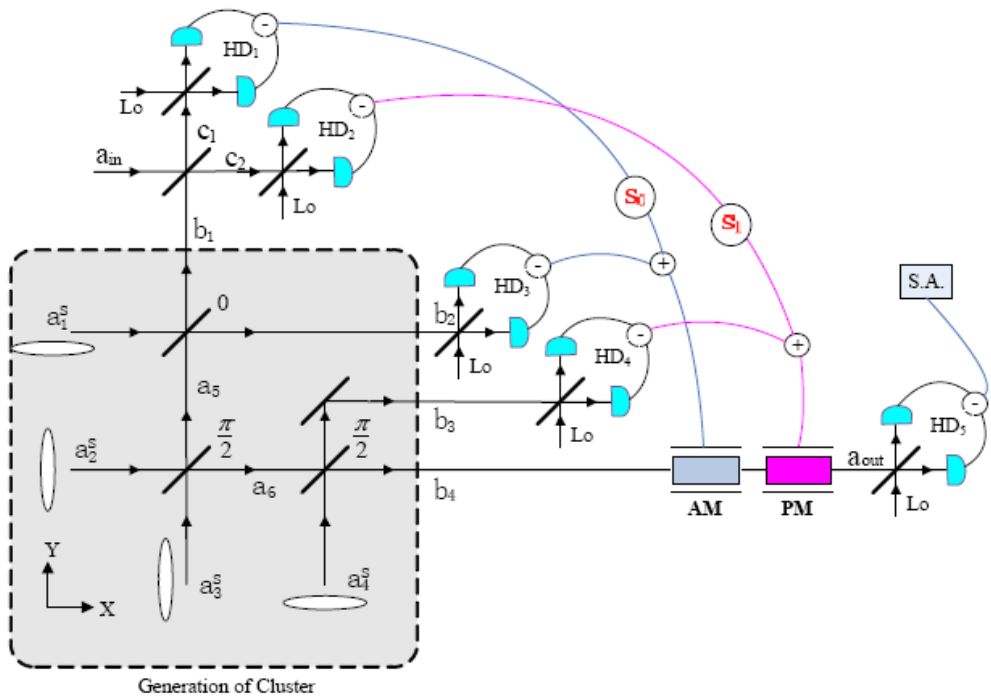

Fig.2 (Color online) The experimental scheme to realize phase-space displacement operation

using quadripartite linear cluster states

PM is a phase modulator；AM is a amplitude modulator；Lo is local oscillator.

$s_0$ and $s_1$ are the values subtracted from the measurement results

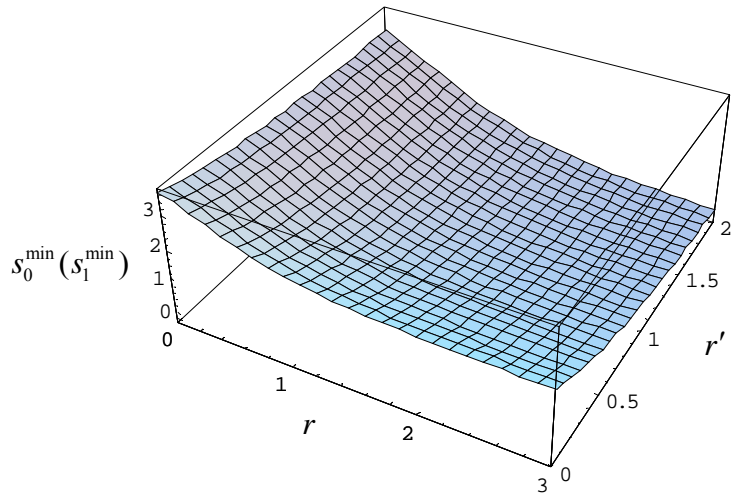

Fig 3 (Color online) Distinguishable displacement $s_0^{min}(s_1^{min})$ vs the initial squeezing parameter $r$ of the Cluster state and the squeeze parameter $r'$ of the input state

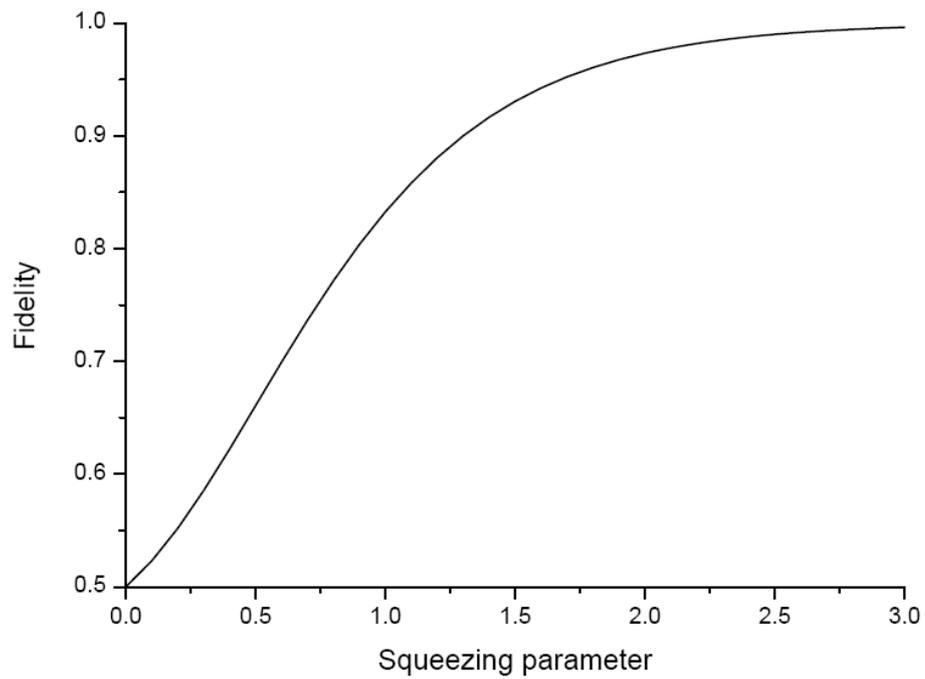

Fig. 4 The fidelity $F$ vs the initial squeezing parameter $r$

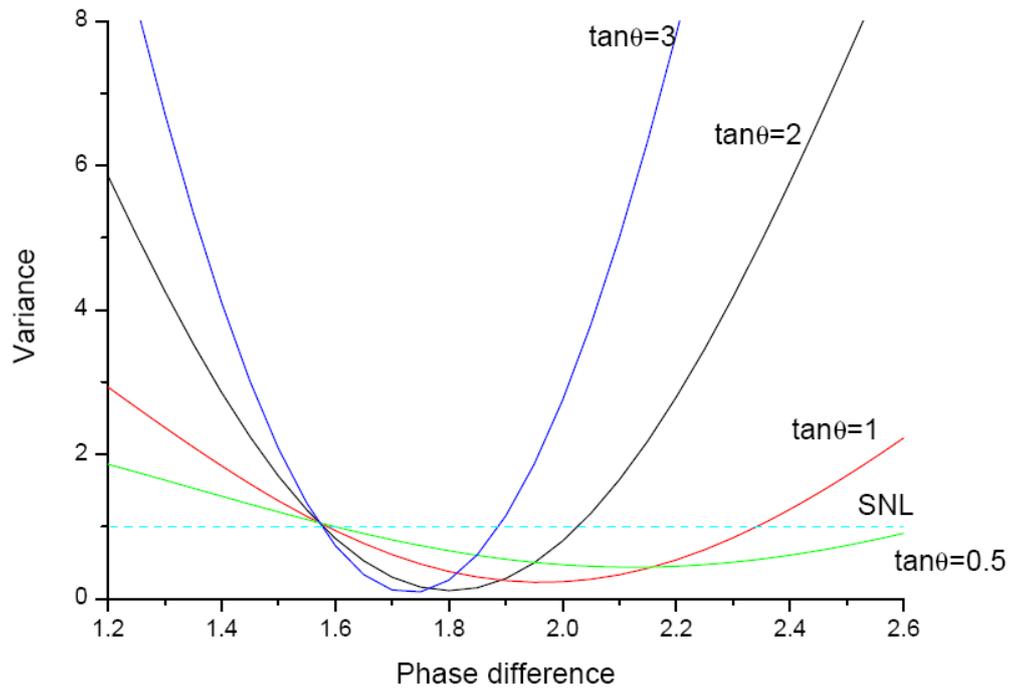

Fig.5 (Color online) Fluctuation variances of the output mode $V(Y^{out}\sin\phi + X^{out}\cos\phi)$

vs phase difference $\phi$ of the HD$_5$ for the different detection phase $\theta$

The dash line is the normalized shot noise limit (SNL), and taking $r = 2$

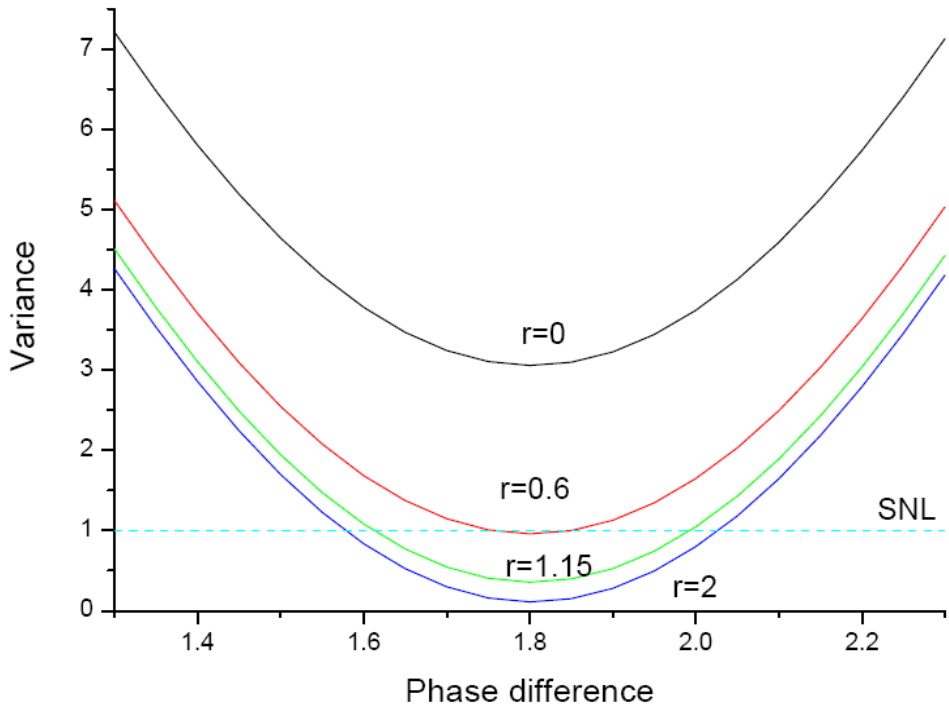

Fig.6 (Color online) Fluctuation variances of the output mode $V(Y^{out}\sin\phi + X^{out}\cos\phi)$

vs phase difference $\phi$ of the HD$_5$ for the different squeezing factor $r$

The dash line is the normalized shot noise limit (SNL), and taking $\tan\theta = 2$

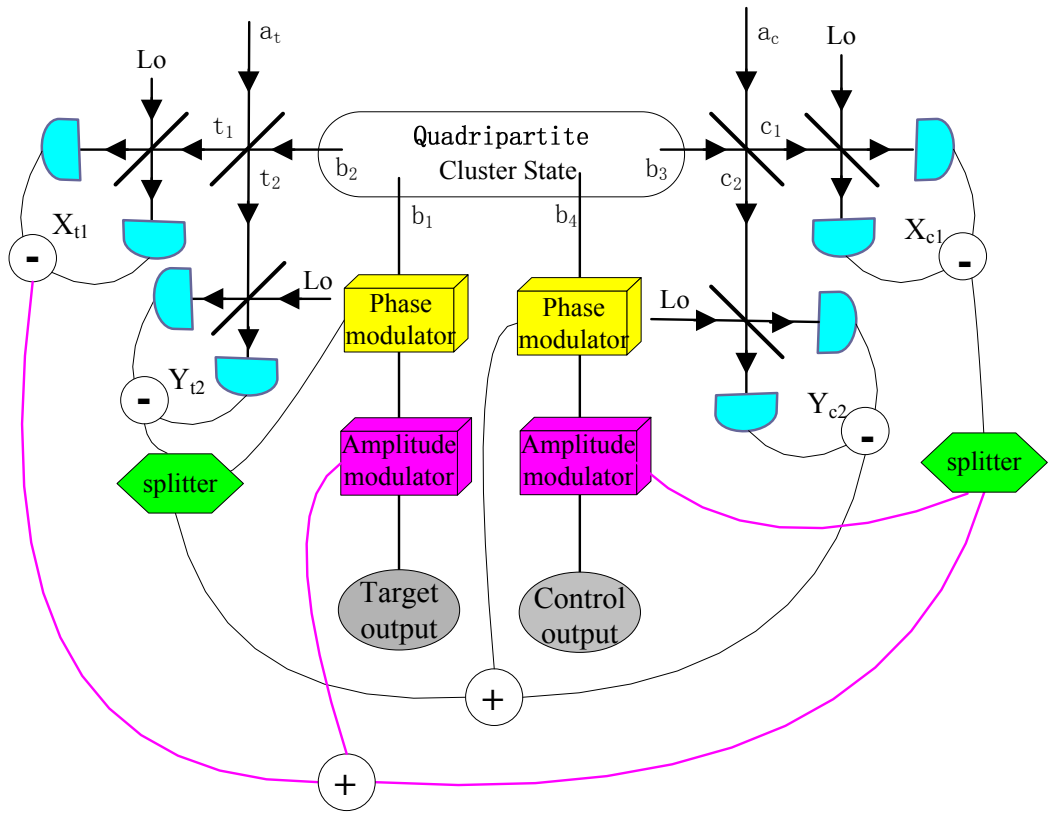

Fig. 7 (Color online)Experiment scheme for realizing CV controlled-X operation using linear quadripartite Cluster state.

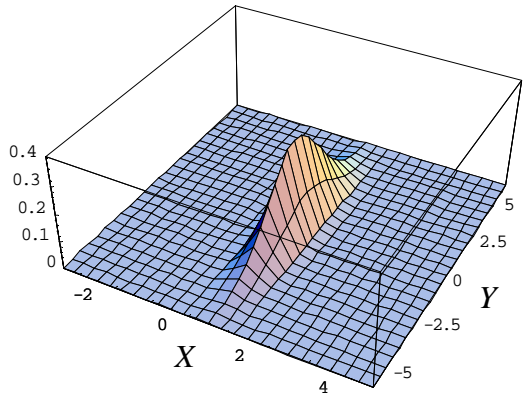 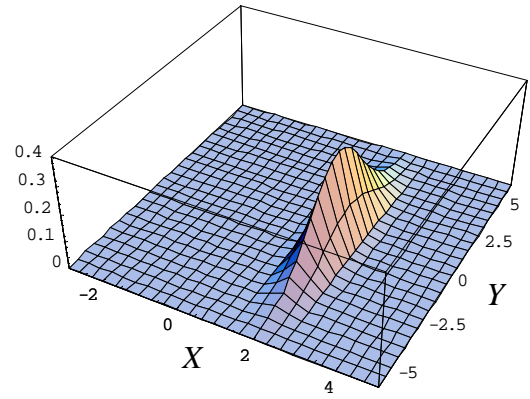

For Control input signal ($\sigma_x = e^{-1}, \sigma_y = e^1$)     For Target input signal ($\sigma_x = e^{-1}, \sigma_y = e^1$)

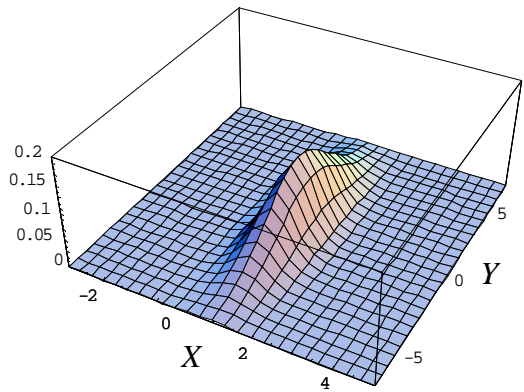 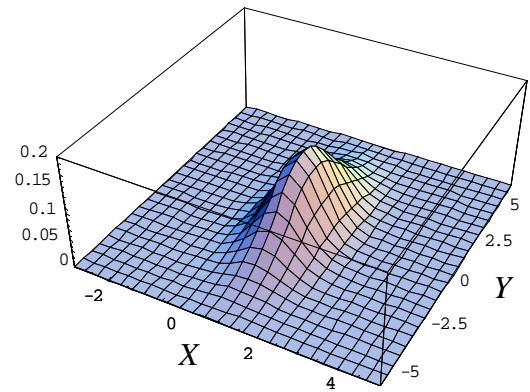

For Control output signal ($r = 1$)     For Target output signal ($r = 1$)

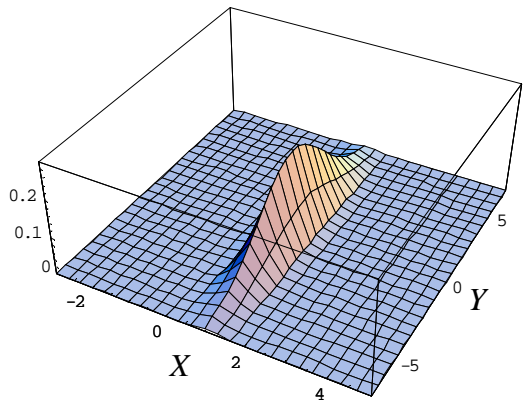 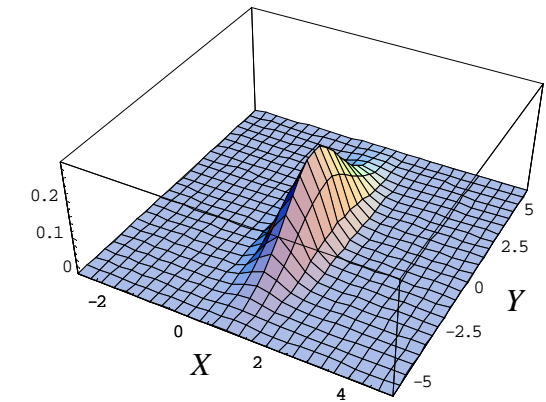

For Control output signal ($r = 3$)     For Target output signal ($r = 3$)

Fig. 8 Wigner functions $W(X,Y)$ of the input signals and output signals

$X$, $Y$ are the amplitude and phase quadrature in the phase space